\documentclass[12pt]{iopart}
\usepackage{graphics}
\usepackage{latexsym}
\begin{document}
\title {{ Protoneutron star in the Relativistic Mean-Field
Theory}}
\author{R. Ma\'{n}ka, M. Zastawny-Kubica, A.Brzezina, I.Bednarek}

\address{Department of Astrophysics and Cosmology,
 Institute of Physics,
 University of Silesia,
 Uniwersytecka 4, 40-007 Katowice,
 Poland.}

\begin{abstract}
In this review the basic properties of nonrotating and slowly
rotating protoneutron stars in the relativistic mean-field
approach are discussed. The equation of state is the main input
to the structure equations. The TM1 parameter set extended to the
case of finite temperature is used to obtain the mass-radius
relation for protoneutron stars. The occurrence of unstable areas
in the mass-radius relation are presented. This allows for the
existence of distinctively different evolutionary of tracks of
the protoneutron stars. The low density protoneutron star
configurations are estimated. The obtained stable configurations
for the fixed lepton number $Y_{L}=0.4$ are compared with ones
obtained for the fixed proton fraction $Y_{P}=0.1776$.

\end{abstract}


\submitto{\JPG}
\maketitle

\section{Introduction}

Protoneutron star is a hot, lepton-rich neutron object which is
formed in Type-II supernovae explosion as the final stage of a
star more massive than about 8 \( M_{\odot } \) - 25 \( M_{\odot
} \) \cite{kot2}. The evolution of a protoneutron star depends on
two factors: the relativistic equation of state of the stellar
matter and neutrino interactions which provide to the phenomena
of neutrino trapping in a neutron star matter . Neutrino
interactions play a crucial role during the creation of the
protoneutron star. At the beginning, this star is very hot and
lepton-rich, after a typical time of several tens of seconds, the
star becomes cold, deleptonized object and the neutron star is
formed. During the collapse enormous energy of the order of \(
\sim 10^{53} \) ergs is released just through neutrinos. The
released energy is equal to the gravitational binding energy of a
newly formed neutron star. In this paper the mass-radius relation
of the protoneutron star is constructed for different temperature
cases. A particularly interesting situation results in the
occurrence of unstable areas which allows distinctively different
evolutionary tracks. In the new-born protoneutron star (less than
several seconds) neutrinos are trapped locally in the dense
stellar matter at densities greater than  \( 10^{13} \) \(
g/cm^{3} \), forming an ultrarelativistic and degenerate Fermi
gas. This
trapped neutrinos have an influence on the equation of state 
\cite{kot2, kot0, kot1, kot3, kot4}.\\
The outline of this paper is as follows.\\
In Sect.1 the general properties of nonrotating and rotating
protoneutron stars are calculated. In Sect.2 the employed
equation of state is obtained in the approach of relativistic
mean-field approximation. This approach imply the nucleons
interactions through the exchange of meson fields. Thus the model
considered here comprises: electrons, muons, neutrinos and
scalar, vector-isoscalar and vector-isovector mesons.
Consequently the pressure and energy density are characterized by
contributions coming from these components. In Sect.3 the
influence of rotation on the protoneutron stars parameters such
as: masses, radii and moment of inertia are calculated. The
obtained results for the given form of the equation of state
followed by a discussion of the implications of these results are
presented in Sect.4 and
Sect.5.\\
 In this article the signature \( \{-,+,+,+\} \) is
used, in which the Dirac's matrix is defined by
\begin{eqnarray*}
\{\gamma ^{\mu },\gamma ^{\nu }\}=-2g^{\mu \nu }I.
\end{eqnarray*}
 and\begin{eqnarray*}
\Box =\partial _{\mu }\partial ^{\mu }=\triangle -\frac{\partial
^{2}}{\partial t^{2}}
\end{eqnarray*}
and $\hbar =c=1$.
 In case of the fermion fields it is more convenient to use the reper field
\( e^{a}_{\mu } \) defined as follows \( g_{\mu \nu }=e^{a}_{\mu
}e^{b}_{\nu }\eta _{ab} \) where \( \eta _{ab} \) is the flat
Minkowski space-time matrix.
\section{The theoretical model}
This paper presents a basic model of a protoneutron star in the
RMF approximation \cite{osa1,osa2,osa3,osa4,osa5}. The
relativistic approach to high-density nuclear matter in this
approach was proposed by Walecka. Since then, this theory has been
applied successfully to many subjects in nuclear many body
problems. Especially the Walecka model (QHD) and its nonlinear
extensions have been quite successfully and widely used for the
description of hadronic matter and finite nuclei. Increasing
interest in the neutron star matter at finite temperature has
been observed recently in relation to the problems of hot neutron
stars and protoneutron stars and their evolution in particular.
Theories concerning protoneutron stars are being discussed in
works by Prakash et al. \cite{kot2,kot0,kot3,kot6,kot7,kot8}. Glendenning
\cite{osa6} has studied the properties of neutron star in the
framework of nuclear relativistic field theory. In the field
theoretical approach the nuclear matter is described as the
baryons interaction through mesons \( \sigma  \), \( \omega  \)
and \( \rho  \) exchange. We define an action of the protoneutron
star through the Lagrangian density ${\mathcal{L}}$
\begin{eqnarray} S=\int
d^{4}x\sqrt{-g}{\mathcal{L}}.
\end{eqnarray}
 The Lagrangian density function of this theory can be represented as the sum
 \begin{equation}
 \label{lag}
{\mathcal{L}}={\mathcal{L}_{B}}+{\mathcal{L}_{L}}+{\mathcal{L}_{\mathcal{M}}}+{\mathcal{L}_{G}},
\end{equation}
 where \( {\mathcal{L}_{B}} \), \( {\mathcal{L}_{L}} \), \( {\mathcal{L}_{\mathcal{M}}} \),
\( {\mathcal{L}_{G}} \) describe the baryonic, leptonic, mesonic
and gravitional sector, respectively.
 The fermion fields are composed of neutrons, protons, muons, electrons and neutrinos
 \begin{equation}
\psi =\left( \begin{array}{c}
\psi _{p}\\
\psi _{n}
\end{array}\right) ,\, \, L_{i}=\left( \begin{array}{c}
\nu _{e}\\
e^{-}
\end{array}\right) _{L},\, \, L_{L}=\left( \begin{array}{c}
\nu _{\mu }\\
\mu ^{-}
\end{array}\right) _{L},\, \, e_{R}=\left( \begin{array}{cc}
e_{R}^{-},\, \mu _{R}^{-}
\end{array}\right) .
\end{equation}
 The baryonic sector of the Lagrangian density function is given by
 \begin{eqnarray}
{\mathcal{L}_{B}}=i\overline{\psi }\gamma ^{\mu }D_{\mu }\psi -\overline{\psi }(M-g_{\sigma }\varphi )\psi ,
\end{eqnarray}
 where $D_{\mu}$ is the covariant derivative
  \begin{eqnarray}
D_{\mu }=\partial _{\mu }+\frac{1}{2}ig_{\rho }\rho _{\mu }^{a}\sigma ^{a}+ig_{\omega }\omega _{\mu }.
\end{eqnarray}
 The leptonic part of the Lagrangian (\ref{lag}) is define as
 \begin{eqnarray}
 \label{lag1}
{\mathcal{L}_{L}}=i\sum _{f}\bar{L}_{f}\gamma ^{\mu }\partial _{\mu }L_{f}-\sum _{f}g_{f}(\bar{L}_{f}He_{Rf}+h.c),
\end{eqnarray}
 where \( H \) is Higgs field and this field has only the residual form\[
H=\frac{1}{\sqrt{2}}\left( \begin{array}{c}
0\\
V
\end{array}\right) \]
 the value of \( V=250 \) \( GeV \) comes from the electroweak interaction
scale. In the mesonic sector the Lagrangian density function is
given by
\begin{eqnarray*}
{\mathcal{L}_{\mathcal{M}}}=-\frac{1}{2}\partial _{\mu }\varphi \partial ^{\mu }\varphi -U(\varphi )-\frac{1}{4}F_{\mu \nu }F^{\mu \nu }-\frac{1}{4}W_{\mu \nu }W^{\mu \nu }-\frac{1}{2}M_{\omega }^{2}\omega _{\mu }\omega ^{\mu } &  & \\
-\frac{1}{4}c_{3}(\omega _{\mu }\omega ^{\mu })^{2}-\frac{1}{4}R_{\mu \nu }^{a}R^{a\mu \nu }-\frac{1}{2}M_{\rho }^{2}\rho _{\mu }^{a}\rho ^{a\mu }, &
\end{eqnarray*}
 where \( \varphi  \) is the scalar field, \( F_{\mu \nu } \) is the electromagnetic
stress tensor \begin{eqnarray*}
F_{\mu \nu }=\partial _{\mu }A_{\nu }-\partial _{\nu }A_{\mu }
\end{eqnarray*}
and $W_{\mu\nu}$ and $R^{a}_{\mu\nu}$ are vector mesons field
strength given by
\begin{eqnarray*}
W_{\mu \nu }=\partial _{\mu }\omega _{\nu }-\partial _{\nu
}\omega _{\mu },
\end{eqnarray*}
\begin{eqnarray*}
R_{\mu \nu }^{a}=\partial _{\mu }\rho _{\nu }^{a}-\partial _{\nu
}\rho _{\mu }^{a}+g_{\rho }\varepsilon _{abc}\rho _{\mu }^{b}\rho
_{\nu }^{c}
\end{eqnarray*}
 \( M_{\omega } \) and \( M_{\rho } \)
are the meson \( \omega  \) and \( \rho  \) masses.
 The potential function $U(\varphi )$ possesses a polynomial form
 \begin{eqnarray*}
U(\varphi )=\frac{1}{2}M_{\sigma }^{2}\varphi
^{2}+\frac{1}{3}g_{2}\varphi ^{3}+\frac{1}{4}g_{3}\varphi ^{4}
\end{eqnarray*}
 introduced by Boguta and Bodmer \cite{Bogut77} in order to get a correct
 value of the compressibility  $K$ of nuclear matter at saturation
 density.
The model described by the Lagrangian function (\ref{lag}) with
such an ansatz  for the potential $U(\varphi )$ guarantee a very
good description of bulk nuclear matter properties for different
parameter sets. It is not possible for neutron star matter to be
purely neutron one. As it was stated above the fermion fields are
composed of protons, neutrons, electrons, muons and neutrinos
because we are dealing here with the electrically neutral neutron
star matter being in $\beta$-equilibrium. Such a matter possesses
a highly asymmetric character caused by the presence of small
amounts of protons and electrons. Finally, the standard
gravitational sector has the form
\begin{eqnarray} {\mathcal{L}_{G}}=\frac{1}{2\kappa }R
\end{eqnarray}
 where \( {\kappa }=8 \pi G /c^{4}\).
 The parameters entering the Lagrangian function  are the coupling
 constants $g_{\omega}$, $g_{\rho}$ and $g_{\sigma}$ and
 self-interacting coupling constants $g_{2}$, $g_{3}$. The
 parameters employed in this model are collected
in the Table \ref{tab: 1}. In our calculations, we used the TM1
\cite{osa7} parameter set, which has a capability to reproduce
the known results of finite nuclei as well as normal nuclear
matter.
\begin{table}
\begin{tabular}{|c|c|c|c|c|}
\hline
\( g_{\sigma }\, \,  \)&
 \( g_{\omega }\, \,  \)&
 \( g_{\rho }\, \,  \)&
 \( g_{2}[\, MeV\, ] \)&
 \( g_{3} \)\\
\hline
\( 10.0289 \)&
 \( 12.6139 \)&
 \( 4.6322 \)&
 \( 1427.18 \)&
 \( 0.6183 \)\\
\hline \( M_{\sigma }\, [\, MeV\, ] \)&
 \( M_{\omega }\, [\, MeV\, ] \)&
 \( M_{\rho }\, [\, MeV\, ] \)&
 \( M\, [\, MeV\, ] \)&
 \( c_{3} \)\\
\hline
\( 511.198 \)&
 \( 783 \)&
 \( 770 \)&
 \( 938 \)&
 \( 71.3075 \) \\
\hline
\end{tabular}
\caption{\label{tab: 1}The parameters set of the model \cite{osa7}.}
\end{table}
 The equations of motion obtained from the Lagrangian function
 for individual fields  are the Dirac equations for nucleon
 fields $\psi$
 \begin{equation}
\label{eq1} i\gamma ^{\mu }D_{\mu }\psi -m_{F}\psi =0
\end{equation}
with $m_{F}$ being the effective nucleon mass
\begin{eqnarray}
m_{F}=M\, \delta =M-g_{\sigma }\varphi
\end{eqnarray}
and Klein-Gordon equations with source terms for meson fields
$\varphi$, $\omega_{\mu}$, $\rho_{\mu}^{a}$
 \begin{eqnarray}
\Box \varphi =M_{\sigma }^{2}\varphi +g_{2}\varphi
^{2}+g_{3}\varphi ^{3}-g_{\sigma }\overline{\psi }\psi \label{eq0}
\end{eqnarray}
\begin{equation}
\label{eq2}
\partial _{\mu }W^{\mu \nu }=M_{\omega }^{2}\omega ^{\nu }+c_{3}(\omega _{\mu }\omega ^{\mu })\omega ^{\nu }+g_{\omega }J^{\nu
}_{B}
\end{equation}
\begin{equation}
\label{eq3} D_{\mu }R^{\mu \nu a}=M_{\rho }^{2}\rho ^{\nu
a}+g_{\rho }J^{a\nu }.
\end{equation}
 Sources that appear in the equations of motion are the baryon
 current
 \begin{equation}
J^{\nu }_{B}=\overline{\psi }\gamma ^{\nu }\psi
\end{equation}
 and existing only in the asymmetric matter the isospin current
 \begin{equation}
J^{a\nu }=\overline{\psi }\gamma ^{\nu }\frac{1}{2}\sigma ^{a}\psi .
\end{equation}
 The baryon and isospin charges are defined by the zero component of the adequate currents,
so\begin{eqnarray}
Q_{B}=\int d^{3}x\, n_{B}=\int d^{3}x\, J^{0}=\int d^{3}x\psi ^{+}\psi =\int d^{3}x(\psi ^{+}_{p}\psi _{p}+\psi ^{+}_{n}\psi _{n}) &
\end{eqnarray}
 and \begin{eqnarray}
Q_{3}=\int d^{3}x\overline{\psi }\gamma ^{0}\frac{1}{2}\sigma ^{3}\psi =\frac{1}{2}\int d^{3}x(\psi ^{+}_{p}\psi _{p}-\psi ^{+}_{n}\psi _{n}).
\end{eqnarray}
 The form of the energy-momentum tensor $T_{\mu\nu}$ that results
 from the variational principle is given by:
 \begin{eqnarray}
T_{\mu \nu }=2\frac{\partial {\mathcal{L}_{\mathcal{M}}}}{\partial
g^{\mu \nu }}+e_{\mu }^{a}\frac{\partial
{\mathcal{L}_{F}}}{\partial e^{a\nu }}-g_{\mu \nu
}{\mathcal{L}}_{matt}.
\end{eqnarray}
The aim of this paper is to achieve the equation of state of the
protoneutron star matter at finite temperature. Our calculations
are based on the variational method incorporating the
Feynman-Bogoliubov inequality. They were presented in details in
paper \cite{osa5}.
 The total pressure of the protoneutron star appearing in
\( T_{\mu \nu } \)
 is the sum of fermion and meson parts $P=P_{F}+P_{\mathcal{M}}$,
\begin{eqnarray}
P=P_{F}+\frac{1}{2}\left(\frac{M_{\omega}}{M}\right)^{2} \omega
^{2}+\frac{1}{4}c_{3}\omega
^{4}+\frac{1}{2}g_{\rho}^{2}\left(\frac{M}{M_{\rho}}\right)^{2}Q_{3}^{2}-U(\varphi).
\end{eqnarray}
 Whereas the total energy density coming from fermion and meson
contributions $\varepsilon =\varepsilon _{F}+\varepsilon
_{\mathcal{M}}$
  can be written as
 \begin{eqnarray}
\varepsilon = \varepsilon
_{F}-\frac{1}{2}\left(\frac{M_{\omega}}{M}\right)^{2} \omega
^{2}-\frac{1}{4}c_{3}\omega^{4}+g_{\omega}\omega
Q_{B}+\frac{1}{2}g_{\rho}^{2}\left(\frac{M}{M_{\rho}}\right)^{2}Q_{3}^{2}+U(\varphi).
 \end{eqnarray}
 The fermion pressure and the energy density in the case of massive particles
 are defined as
 \begin{eqnarray*}
P_{F}=\sum_{i}\frac{1}{3\pi^{2}}\int_{0}^{\infty}dk
k^{2}\frac{k^{2}}{\sqrt{k^{2}+m_{F,i}^{2}}}\left(
f_{i}+\overline{f}_{i}\right),
\end{eqnarray*}
 \begin{eqnarray*}
 \varepsilon _{F}= \sum_{i}\frac{1}{\pi^{2}}\int_{0}^{\infty}dk
k^{2}\sqrt{k^{2}+m_{F,i}^{2}}(f_{i}+\overline{f}_{i})
\end{eqnarray*}
 where $f_{i}$ and $\overline{f_{i}}$ are the Dirac fermion distribution
 functions.
The pressure $(P_{F})$ and the energy density $(\varepsilon_{F})$
can be written with the use of the functions $\phi_{i}$ and
$\chi_{i}$ as
\begin{eqnarray}
\label{cis}
 P_{F,i}=P_{0,i}\phi_{i}(x_{i}),
\end{eqnarray}
\begin{eqnarray*}
\varepsilon_{F,i}=\varepsilon_{0,i}\chi_{i}(x_{i})
\end{eqnarray*}
where
\begin{eqnarray*}
P_{0,i}=\varepsilon_{0,i}=\frac{m_{F,i}c^{2}}{\lambda_{i}^{3}},\, \,
\, \,i=p,n,e
\end{eqnarray*}
$\lambda$ is the Compton wavelength. The dimensionless functions
$\phi_{i}$ and $\chi_{i}$ are stated in the following way
\begin{eqnarray}
& \phi _{i}(z,t)=\frac{1}{3\pi ^{2}}\int _{0}^{\infty
}\frac{x^{4}dx}{\sqrt{x^{2}+\delta _{i}^{2}}}\left\{
f_{i}(x,z,t)+\overline{f}_{i}(x,z,t)\right\} =& \label{row} \\
& =\frac{1}{3\pi ^{2}}(\frac{\lambda_{i}}{\lambda
_{T}})^{4}H_{5}(r,y_{i}) & \nonumber
\end{eqnarray}
\begin{eqnarray}
 & \chi _{i}(z,t)=\frac{1}{\pi ^{2}}\int _{0}^{\infty }x^{2}dx\sqrt{x^{2}+\delta _{i}^{2}}\left\{ f_{i}(x,z,t)+\overline{f}_{i}(x,z,t)\right\} = & \label{f2} \\
 & =\frac{1}{\pi ^{2}}(H_{5}(r,y_{i})+m_{i}^{2}H_{3}(r,y_{i})) & \nonumber
\end{eqnarray}
with the distribution functions
\begin{equation}
\label{fi3} f_{i}(x,z,t)=\frac{1}{e^{\left( \sqrt{x^{2}+\delta
_{i}^{2}}-z\right) \slash t}+1},
\end{equation}
\begin{equation}
\label{fi4} \overline{f}_{i}(x,z,t)=\frac{1}{e^{\left(
\sqrt{x^{2}+\delta _{i}^{2}}+z\right) \slash t}+1}.
\end{equation}
 In those relations $ m_{F,i}=M \delta _{i}$, $ x= k_{F} /
{M} $, $ z=\mu /M$, $r=z/t=\mu/k_{B}T$, $ y_{i}=\delta_{i}
/t=m_{F,i}/k_{B}T$ and $ t=k_{B}T/M$ are the effective mass, the
dimensionless Fermion momentum, chemical potential and
temperature, respectively. \( \lambda _{T}=1 /k_{B}T \) is the
thermal wavelength ($\delta_{i}=\delta$ for $i=p,n$ end $\delta_{e}=m_{e}/M$
where $M$ is the neutron mass).
\begin{figure}
{\par\centering \resizebox*{10cm}{!}{\includegraphics{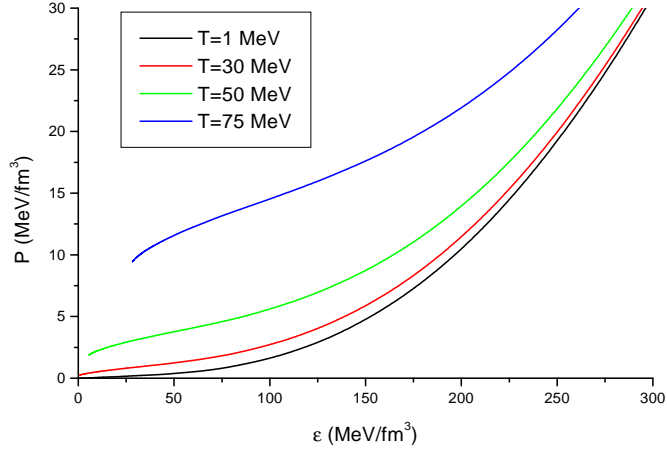}}
\par} \caption{The equation of state for different temperature cases. }
\end{figure}
The thermodynamic properties of fermions can be evaluated using
the well know functions $H_{n}$ and $G_{n}$ given by
\begin{eqnarray}
\label{ener}
H_{n}=\frac{1}{3\pi^{2}}\int_{0}^{\infty}\frac{k^{n-1}dk}{\sqrt{k^{2}+m_{F}^{2}}}\left\{f+\overline{f}\right\}
\end{eqnarray}
\begin{eqnarray}
\label{ener1}
G_{n}=\frac{1}{\pi^{2}}\int_{0}^{\infty}k^{n-1}dk\left\{f-\overline{f}\right\}
\end{eqnarray}
The two terms in (\ref{ener},\ref{ener1}) correspond to the
contributions
of particles and antiparticles, respectively. Our first step is
to calculate the pressure and energy density for nucleons which
correspond to the nonrelativistic case. The relativistic case,
corresponding to the high temperature approximation, is utilized
in order to calculate the pressure coming from electrons and
massive neutrinos. In this approach the function $H_{5}$ and
$H_{3}$ are given by the following relations:
\begin{equation}
\label{h5a} H_{3}(r,y)=\frac{1}{8}y^{2}(-1+2r^{2}+2\Gamma
)+\frac{1}{4}y^{2}ln(y/\pi )
\end{equation}
 and \begin{equation}
\label{h5b}
H_{5}(r,y)=-\frac{1}{32}r^{2}y^{2}+\frac{1}{96}(ry)^{4}-\frac{1}{64}y^{4}ln(y/\pi
)+\frac{3}{256}-\frac{1}{64}\Gamma y^{4}.
\end{equation}
 In the case of massless neutrino (\( m_{\nu}=0 \)) the
 pressure $P$ and the energy density $\varepsilon$ can be expressed
 analogously to the massive fermion relation (\ref{cis}). In this
 situation
 \begin{eqnarray*}
P_{0\nu}=k_{B}^{4}T^{4}
 \end{eqnarray*}
 and function $\phi$ is given by
\begin{equation} \label{f0a} \phi _{0}(z,t)=\frac{1}{3}\chi
_{0}(z,t)=\frac{1}{3\pi ^{2}}\int _{0}^{\infty }x^{3}dx\left\{
\frac{1}{e^{(x-z)\slash t}+1}+\frac{1}{e^{(x+z)\slash
t}+1}\right\} .
\end{equation}
The function $\phi$ can be written with the use of the function
$H_{n}$ defined as
\begin{eqnarray*}
H_{n}(r,y=0)=\int _{0}^{\infty }x^{n-2}dx\left\{
\frac{1}{e^{(x-z)\slash t}+1}+\frac{1}{e^{(x+z)\slash
t}+1}\right\}
\end{eqnarray*}
In agreement with previous substitution the variable
$r=z/t=\mu/k_{B}T$ and $z=\mu/M$. In this moment the case $y=0$
is considered.
 Having integrated these functions (\ref{f0a}) one can obtain the results in
 which the solution is expressed by the polylogaritmic function $Li_{n}(z)$
 \begin{equation}
\label{f0b}
\phi _{0}(z,t)=-\frac{6}{\pi ^{2}}(Li_{4}(-e^{-z/t})+Li_{4}(-e^{z/t})).
\end{equation}
 In the case when the temperature ($T=0$) the integral $\phi _{i}$  and $\chi _{i}$
 give as a solution the well-known
 Shapiro resalt \cite{kot5}:
\begin{eqnarray*}
 \phi(x)=\frac{1}{8 \pi ^2} \left\{x
\sqrt{1+x^2} \left( \frac{2 x^2}{3}-1 \right)+ \ln(x+
\sqrt{1+x^2}) \right\},
\end{eqnarray*}
\begin{eqnarray*}
 \chi(x)=\frac{1}{8 \pi ^2} \left\{x
\sqrt{1+x^2} (1+2x^2)-\ln(x+ \sqrt{1+x^2}) \right\}.
\end{eqnarray*}
 The total \( \phi \) function is sum
of the partial components
\begin{eqnarray*} \phi =\phi _{n}+\phi _{p}+\phi _{e}+\phi _{\nu
}.
\end{eqnarray*}
 The protoneutron star can be characterized by two parameters, the lepton fraction
\( Y_{L} \) and the entropy per nucleon \( S\). The fermion number
density and the entropy per nucleon are giving by the relations
\begin{eqnarray}
\label{licz}
n_{i}=\frac{1}{\pi ^{2}}\int _{0}^{\infty
}x^{2}dx\left\{ f_{i}(x,z,t)-\overline{f}_{i}(x,z,t)\right\},
\end{eqnarray}
\begin{eqnarray*}
S\equiv \frac{s}{n_{B}}
\end{eqnarray*}
while the fermions entropy density \( s \) is defined as
\begin{eqnarray}
 & s=\sum _{i}\frac{1}{\pi ^{2}}\int _{0}^{\infty }dx\, x^{2}[-f_{i}(x,z,t)\ln f_{i}(x,z,t)-(1-f_{i}(x,z,t))\ln (1-f_{i}(x,z,t)) & \nonumber \\
 & -\overline{f}_{i}(x,z,t)\ln \overline{f}_{i}(x,z,t)-(1-\overline{f}_{i}(x,z,t))\ln (1-\overline{f}_{i}(x,z,t))] & \nonumber \label{ent}
\end{eqnarray}
 \cite{as4} with the use of the previously defined Dirac distribution functions \( f_{i} \)
and \( \overline{f}_{i} \). Considering the case of massless
neutrinos the equation (\ref{licz}) can be solved analytically and
the final form of the neutrino number density is giving by
 \begin{eqnarray*}
 & n_{0}=\frac{x^{2}t}{2\pi ^{2}}\ln ((1+e^{(x+\mu )/t})/(1+e^{(x-\mu )/t}))+ & \\
 &  & \\
 & \frac{x\, t}{\pi ^{2}}\sum _{p=(-,+)}Li_{2}(-e^{(x+p\mu )/t})+\frac{t^{2}}{\pi ^{2}}\sum _{p=(-,+)}Li_{3}(-e^{(x+p\mu )/t}) &
\end{eqnarray*}
The obtained result allows us to determine the lepton fraction $Y_{L}$, defined as
 \begin{equation}
Y_{L}=\frac{n_{e}+n_{\nu }}{n_{B}}\, \, ,
\end{equation}
 where \( n_{B} \), \( n_{e} \) and \( n_{\nu } \) are the nucleon, electron
and neutrino number densities, respectively. The proton fraction
can be defined as
\begin{eqnarray}
Y_{P}=\frac{n_{p}}{n_{B}},
\end{eqnarray}
where $n_{p}$ is proton number density.
 The difference
between an ordinary neutron star and a protoneutron star is
caused mainly by the high lepton number of the protoneutron star
matter. This is because a star core is opaque to neutrinos. The
lepton number is approximately constant after core bounce \(
Y_{L} \) \( \simeq  \) \( 0.4 \) for the densities above $10^{12}$
 $g/cm^{3}$ \cite{kot2} - \( 10^{13} \) \( g/cm^{3} \)
\cite{kot3} where the neutrinos are trapped. At the time the
entropy per baryon is approximately constant \( S\sim 1 \). Below
this density the neutrinos are free and they can escape. After
the leak out of neutrinos their chemical potential vanishes.
\begin{figure}
{\par\centering
\resizebox*{10cm}{!}{\includegraphics{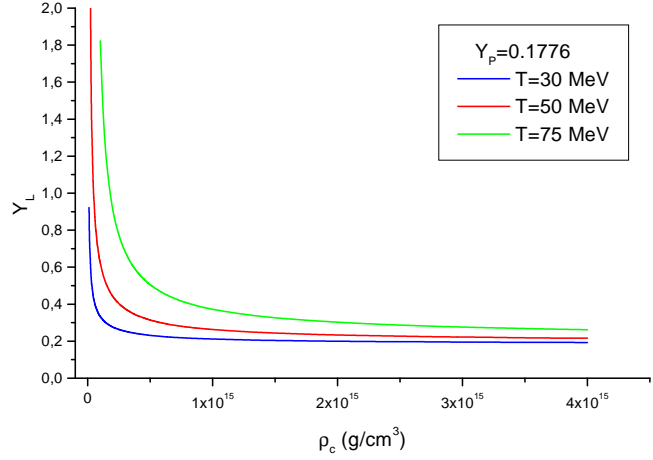}}
\par}
 \caption{The lepton number \protect\( Y_{L}\protect \)
dependence on the central density \protect\( \rho _{c}\protect \)
for the different temperatures and for the constant proton
fraction $Y_{P}=0.1776$.}
\end{figure}
\begin{figure}
{\par\centering \resizebox*{10cm}{!}{\includegraphics{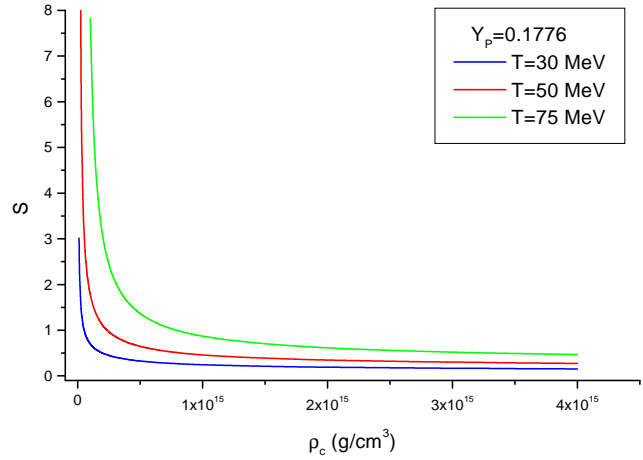}}
\par}
 \caption{The entropy \protect\( S\protect \) dependence on
the central density \protect\( \rho _{c}\protect \) for the
different temperatures and \protect\( Y_{P}=0.1776\protect \).}
\end{figure}
\begin{figure}
{\par\centering
\resizebox*{10cm}{!}{\includegraphics{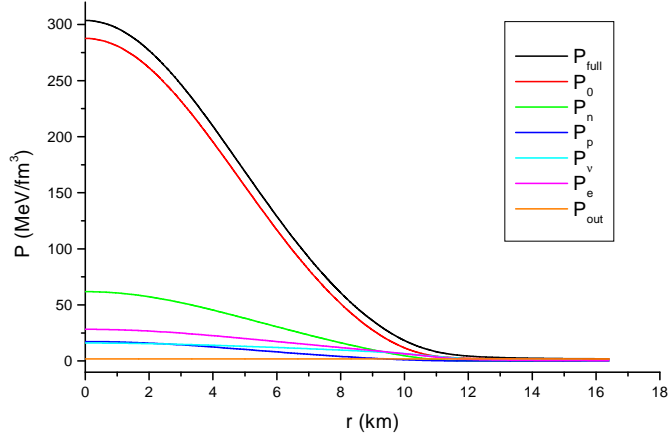}} \par}
\caption{The pressure \protect\( P\protect \) versus the radius
\protect\( r\protect \) for different components of pressure. The
temperature is fixed \protect\( T=50\protect \) \protect\(
MeV\protect \), \protect\( \rho _{c}=1.8\protect \) \protect\(
10^{15}\protect \) \protect\( g/cm^{3}\protect \). }
\end{figure}
\begin{figure}
{\par\centering
\resizebox*{10cm}{!}{\includegraphics{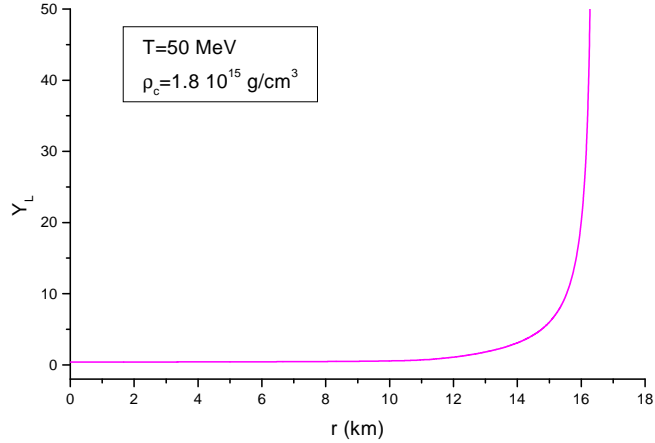}}
\par}
\caption{The lepton number \protect\( Y_{L}\protect \) dependence
on the radius \protect\( r\protect \) for the temperature
\protect\( T=50\protect \) \protect\( MeV\protect \) and
\protect\( \rho _{c}=1.8\protect \) \protect\(
10^{15}g/cm^{3}\protect \).}
\end{figure}
\begin{figure}
{\par\centering \resizebox*{10cm}{!}{\includegraphics{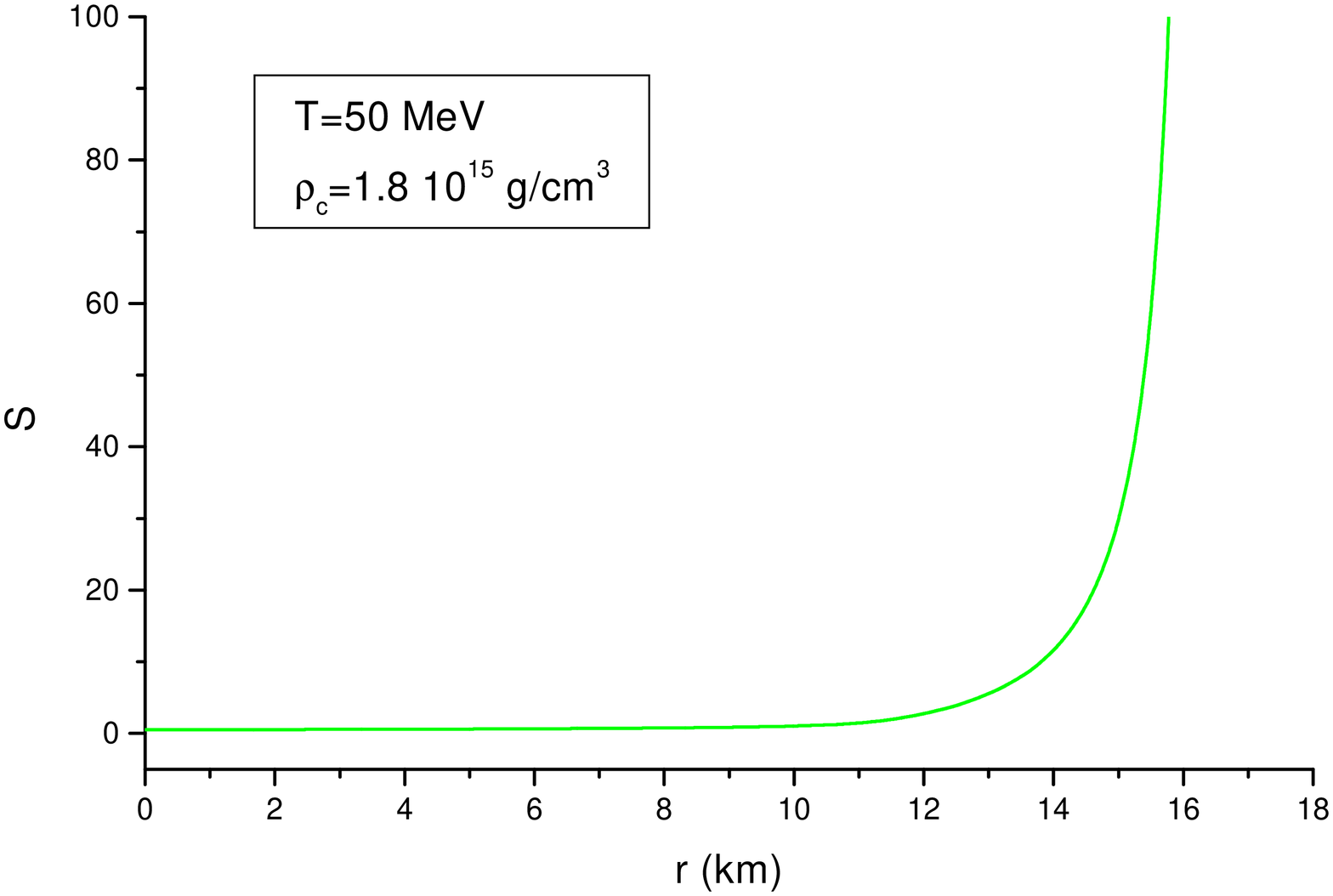}}
\par}
\caption{The entropy \protect\( S\protect \) dependence on the
radius \protect\( r\protect \) for the temperature \protect\(
T=50\protect \) \protect\( MeV\protect \) and \protect\( \rho
_{c}=1.8\protect \) \protect\( 10^{15}g/cm^{3}\protect \).}
\end{figure}
The situation, when the neutrinos are trapped, is characteristic to very initial
stage of a protoneutron star existence. The matter is assumed to be composed
of nucleons and leptons. Thus we are dealing here with electrically neutral matter being in
$\beta$ equilibrium which can be expressed as a relation between the chemical potentials
of the protoneutron star components.
\begin{eqnarray}
\mu _{n}+\mu _{e}=\mu _{\nu _{e}}+\mu _{p}.
\end{eqnarray}
This equation can be written in terms of neutron $x_{n}$ and
proton $x_{p}$ Fermi momentum
\[
x_{p}=\left(\frac{Y_{P}}{1-Y_{P}}\right)^{\frac{1}{3}} x_{n}
\]
 \[
\sqrt{\delta ^{2}_{e}+x_{e}^{2}}+\sqrt{\delta
^{2}_{p}+x_{p}^{2}}=\sqrt{\delta ^{2}_{n}+x_{n}^{2}}+\sqrt{\delta
^{2}_{\nu }+x_{\nu }^{2}}\] for electrons and neutrinos
$(i=e,\nu)$ $\delta_{i}=m_{e}/M,0$ whereas for nucleons
$\delta=\delta_{n}=\delta_{p}$. Fermi momentum can be effectively
only a function of the neutron Fermi momentum \( x_{n} \) and the
\( Y_{P} \) parameter, which measures the proton admixture in the
neutron star. The protoneutron star matter in general is more
symmetric than neutron star matter.\\
\begin{figure}
{\par\centering
\resizebox*{10cm}{!}{\includegraphics{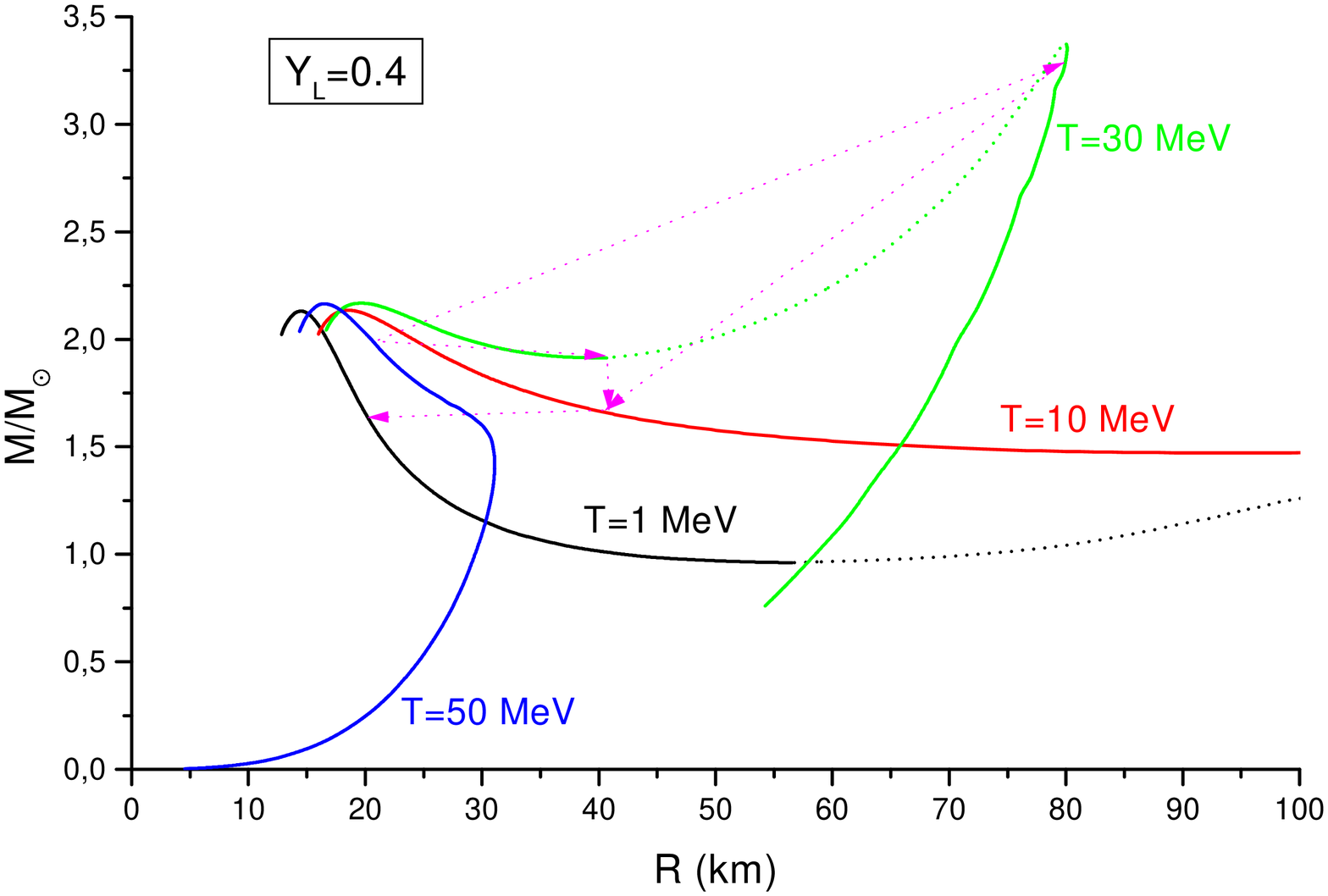}} \par}
\caption{The mass-radius diagram for the protoneutron star for
constant \protect\( Y_{L}=0.4\protect \).}
\end{figure}
\begin{figure}
{\par\centering
\resizebox*{10cm}{!}{\includegraphics{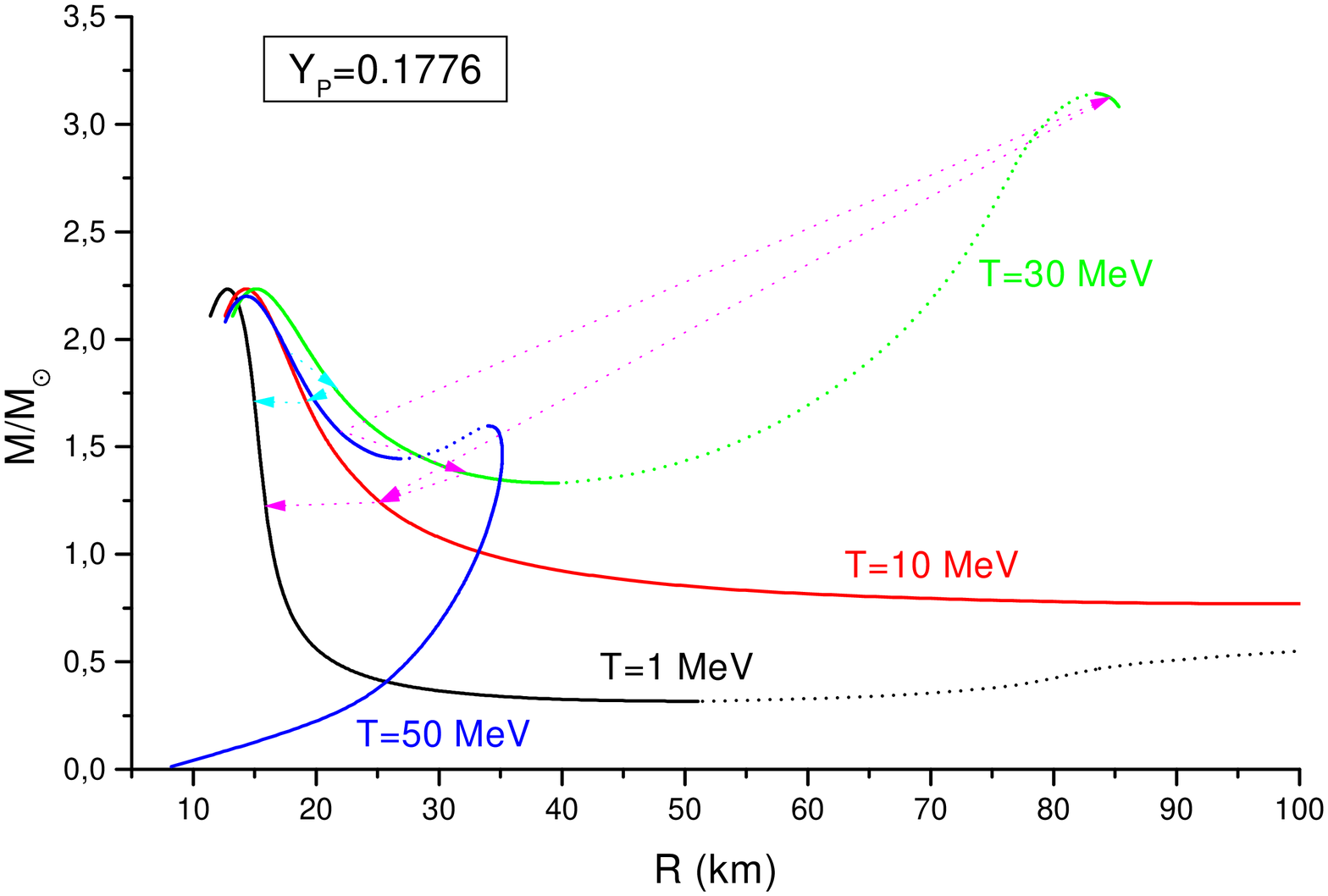}} \par}
\caption{The mass-radius diagram for the protoneutron star for
constant \protect\(Y_{P}=0.1776\protect \).}
\end{figure}
\begin{figure}
{\par\centering \resizebox*{10cm}{!}{\includegraphics{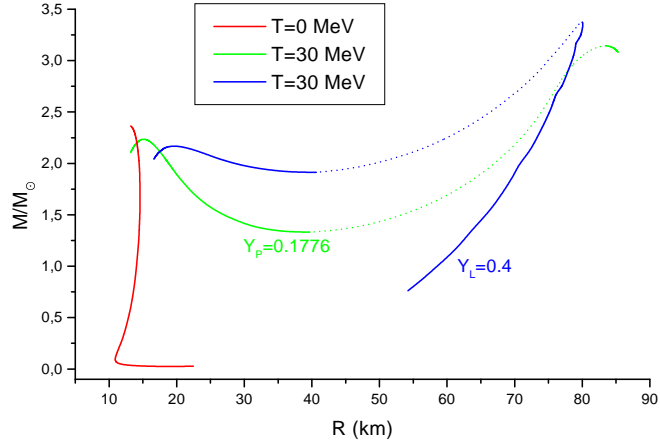}}
\par} \caption{The mass-radius diagram for the protoneutron star with $T=30$ $MeV$
($Y_{L}=0.4$, $Y_{P}=0.1776$) and $T=0$ $MeV$( the proton
fraction is constrained by $\beta$ decay).}
\end{figure}
\begin{figure}
{\par\centering
\resizebox*{10cm}{!}{\includegraphics{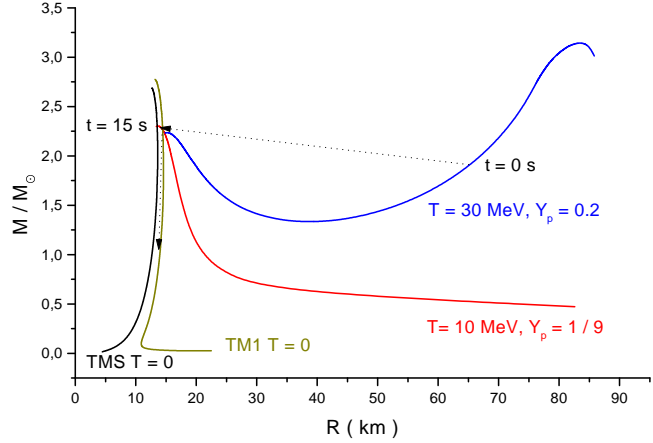}}
\par} \caption{The time evolution ($t=0-15 s$) of the protoneutron star (with $M{B}=1M_{\odot}$)
in accordance with paper Toki \cite{toki}.}
\end{figure}
\begin{figure}
{\par\centering
\resizebox*{10cm}{!}{\includegraphics{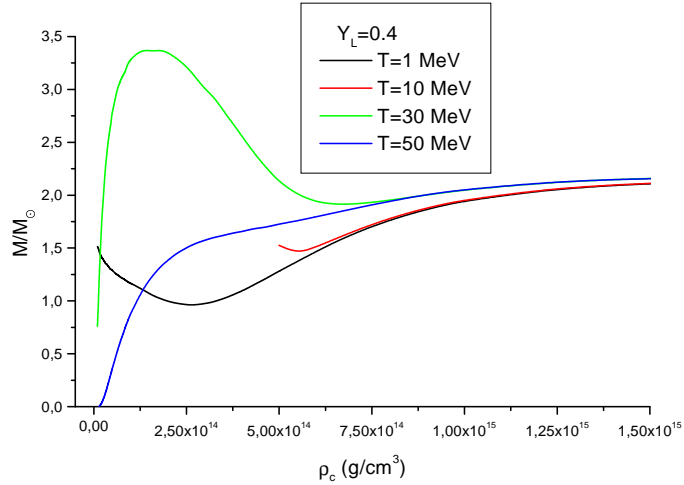}} \par}
\caption{The mass \protect\( M\protect \) dependence on the
central density \protect\( \rho _{c}\protect \) for \protect\(
Y_{L}=0.4\protect \).}
\end{figure}
\begin{figure}
{\par\centering
\resizebox*{10cm}{!}{\includegraphics{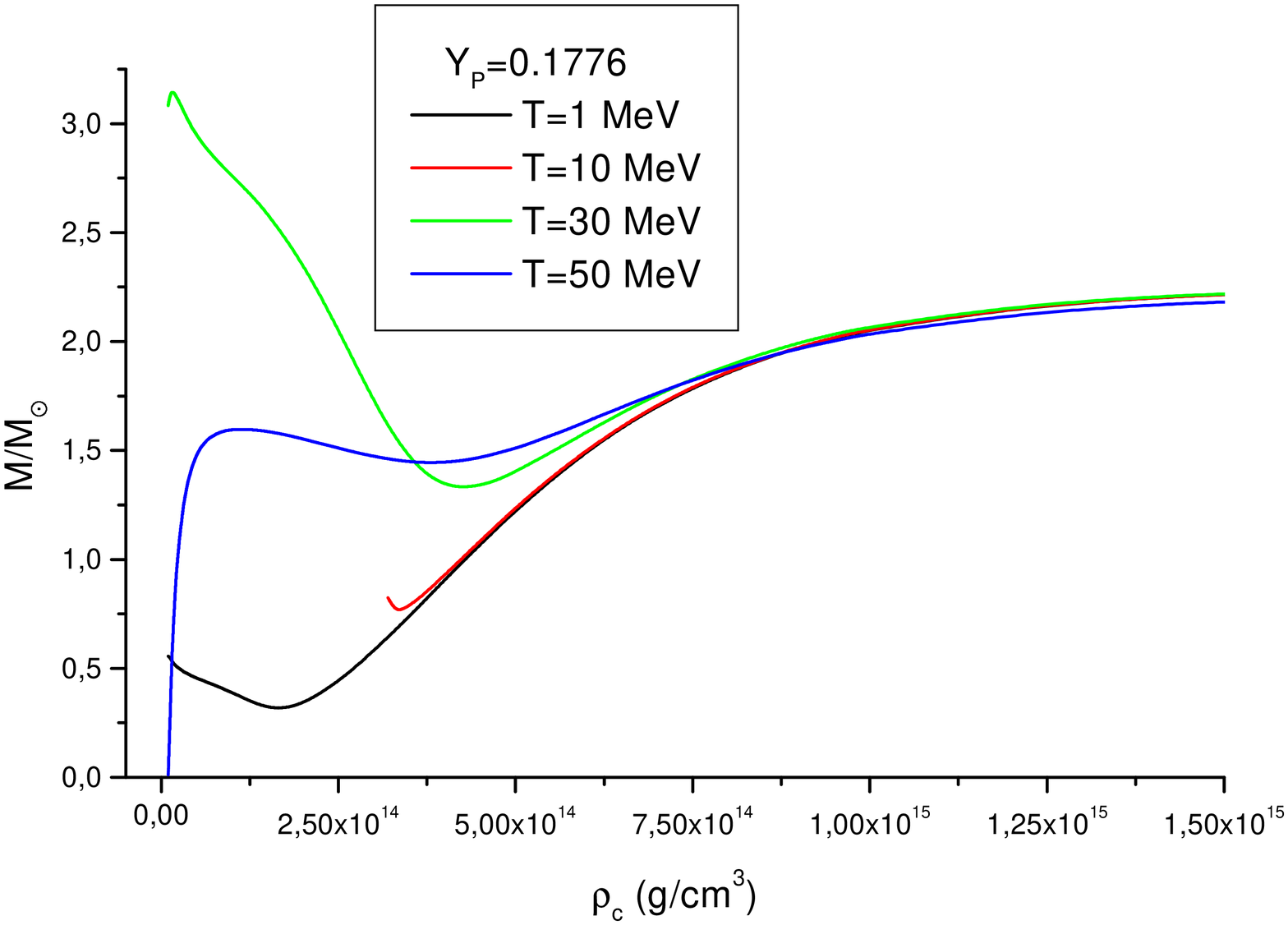}} \par}
\caption{The mass \protect\( M\protect \) dependence on the
central density \protect\( \rho _{c}\protect \) for
$Y_{P}=0.1776$.}
\end{figure}
\section{Rotating protoneutron star}
The spacetime
outside a rotating protoneutron star
is much more complicated than the metric outside a non-rotating
star thus it seems to be  interesting to investigate the properties of rotating protoneutron stars.
In general the metric of a stationary, axisymmetric, asymptotically
flat spacetime has the form
\begin{eqnarray}
\label{metryka1}
 ds^{2}=g_{\mu \nu }dx^{\mu }dx^{\nu }=-e^{2{\nu
}}c^{2}dt^{2}+e^{2{\psi }}(d{\phi }-{\omega }cdt)^{2}+e^{2{\mu
}}d{\theta }^{2}+e^{2{\lambda }}dr^{2},
\end{eqnarray}
with $g_{\mu \nu}$ being the metric tensor. The metric potential
functions $\nu,\psi,\mu$ and $\lambda $ and the angular velocity
$\omega $ of the stellar fluid  in the local inertial frame are
functions of the radial coordinate $r$ and the polar angel
$\theta $ \cite{ak1, ak2}. In order to compute the structure of rapidly rotating
fluid body the numerical method developed by Butterworth and Ipser
\cite{Butter} was introduced. Besides this exact numerical
treatment there is a perturbative Hartle's method which is based
on the assumption that rotating massive body is no longer
spherically symmetric. It is destorted, thus expanding the metric
functions through second order in the stars rotational velocity
$\Omega $ one can obtain the following form of the perturbed
metric
\begin{eqnarray}
& ds^{2}=g_{\mu \nu }dx^{\mu }dx^{\nu }=-e^{2{\nu }(r,\; {\theta
},\; {\Omega })}c^{2}dt^{2}+e^{2{\psi }(r,\; {\theta },\; {\Omega
})}(d{\phi }-{\omega }(r,\; {\theta },\; {\Omega })cdt)^{2}& \nonumber \\
&+e^{2{\mu }(r,\; {\theta },\; {\Omega })}d{\theta
}^{2}+e^{2{\lambda }(r,\; {\theta },\; {\Omega
})}dr^{2}+O(\Omega^{3}),& \nonumber
\end{eqnarray}
where metric functions in this perturbed line element are given by
\begin{eqnarray*}
e^{2{\nu }(r,\; {\theta },\; {\Omega })}=e^{2{\Phi
}(r)}(1+2(h_{0}(r,{\Omega })+h_{2}(r,\; {\Omega
})P_{2}(cos{\theta }))),
\end{eqnarray*}
 \[
e^{2{\psi }(r,\; {\theta },\; {\Omega })}=r^{2}sin^{2}{\theta
}(1+2(v_{2}(r,\; {\Omega })-h_{2}(r,\; {\Omega
}))P_{2}(cos{\theta })),\]
 \[
e^{2{\mu }(r,\; {\theta },\; {\Omega })}=r^{2}(1+2(v_{2}(r,\;
{\Omega })-h_{2}(r,\; {\Omega }))P_{2}(cos{\theta })),\]
\begin{eqnarray}
e^{2{\lambda }(r,\; {\theta },\; {\Omega })}=e^{2{\Lambda
}(r)}\left( 1+\frac{2G}{c^{2}r}\frac{m_{0}(r,\; {\Omega
})+m_{2}(r,\; {\Omega })P_{2}(cos{\theta
})}{(1-\frac{2G}{c^{2}}\frac{m(r)}{r})}\right) ,
\end{eqnarray}
where $\Phi (r)$ and $\Lambda (r)$ are the metric functions of a
spherically symmetric star, $P_2$ the Legendre polynomial of
order 2, $m_0, m_2, h_0, h_2$ and $\upsilon _2$ are all functions
of $r$ and $\Omega $ \cite{as1,osa1}. They are calculated from
Einstein's field equations and given as solutions of Hartle's
stellar structure equations, $\omega $ has the same meaning as in
the nonperturbative line element (\ref{metryka1}). The metric
functions are determined with the use of the Einstein equations
\begin{equation}
G_{\mu \nu }\equiv R_{\mu \nu }-\frac{1}{2}g_{\mu \nu }R={\kappa }T_{\mu \nu },\label{ein1}
\end{equation}
where $T_{\mu \nu }$ is the stress-energy tensor given in the perfect fluid form
\begin{equation}
T_{\mu \nu}=(\epsilon +P)u_{\mu }u_{\nu }+g_{\mu \nu}P
\end{equation}
and \( u_{\mu } \) is the unite four-velocity satisfying the
following condition \begin{eqnarray*} u_{\mu }u^{\mu }=-I.
\end{eqnarray*}
where $\epsilon $ is the total energy density, $P$ the pressure.
Each metric function, namely $\nu, \rho $ and $\omega $ are
functions of the radial coordinate $r$, polar angle $\theta $ and
also the stars angular velocity $\Omega $. In the case of
rotating stars there is an additional dependence of the metric on
the polar angle and the frame dragging frequency $\omega $. The
latter leads to the existence of the non-diagonal term
$g^{t\varphi }$ in the metric tensor. The assumption of uniform
rotation means that the value of $\Omega $ is constant throughout
the star. For uniformly rotating bodies there is a relation
between components of the four-velocity vector $u^{\varphi
}=\Omega u^{t}$. The nonzero components of the four-velocity
vector $u_{\mu}$ of the matter are of the form
\begin{equation}
u^{t}=(e^{2\nu(r,\theta)}-\overline{\omega}^{2}e^{2\psi(r,\theta)})^{-\frac{1}{2}},\hspace{1cm}u^{\varphi
}=\Omega u^{t}
\end{equation}
where
\begin{equation}
\overline{\omega}=\Omega -\omega  \hspace{1cm} \Omega
=\frac{d\varphi }{dt}
\end{equation}
The absolute limit on stable neutron star rotation is the Kepler
frequency $\Omega _K$. It determines the frequency at which the
mass shedding at the stellar equator sets in. The result of the
work of Haensel and Zdunik \cite{haen} shows that the value of the
Kepler frequency can be estimated knowing the value of the mass
and radius of the corresponding nonrotating star and an empirical
relation was given
\begin{equation}
\Omega_{K} \approx
C_{Hz}\sqrt{(M_s/M_{\odot})(R_s/10km)^3}=(0.63-0.67)\times
\Omega_{c}
\end{equation}
where, $C_{Hz}=7700 s^{-1}$ and $\Omega_{c}$ is the Newtonian value and is equal
\begin{equation}
\Omega_{c}=\sqrt{M_s/R_s^3}
\end{equation}
the index $s$ indicates that these values refer to the spherical
configuration.
As a consequence of the perturbative method for the angular velocity of the local inertial frame
appears
\begin{eqnarray}
\frac{d}{dr}\left( r^{4}j(r)\frac{d\overline{\omega }(r)}{dr}\right) +4r^{3}\frac{dj(r)}{dr}\overline{\omega }(r)=0,\label{om_{0}}
\end{eqnarray}
 where \( \overline{\omega }={\Omega }-{\omega } \) and \( j(r)=e^{-{\Phi }(r)}(1-\frac{2G}{c^{2}}\frac{m(r)}{r})^{\frac{1}{2}} \)
 and the boundary conditions are such that \( \overline{\omega }(0)=\overline{\omega }_{c} \)
and \( \left( \frac{d\overline{\omega }}{dr}\right) _{r=0}=0 \).
The angular velocity dependence on the angular velocity in the
center is stated as
\begin{eqnarray} {\Omega }({\omega }_{c})=\overline{\omega
}(R_{s})-\frac{R_{s}}{3}\left( \frac{d\overline{\omega
}(r)}{dr}\right) _{r=R_{s}}.
\end{eqnarray}
 The solution of equation \( (\ref {om_{0}}) \) allows us to determine the
star's momentum of inertia \begin{eqnarray}
I=\frac{J({\Omega })}{{\Omega }}=\frac{8{\pi }}{3c^{2}}\int _{0}^{R_{s}}drr^{4}\frac{{\varepsilon }(r)+P(r)}{(1-\frac{2G}{c^{2}}\frac{m(r)}{r})^{\frac{1}{2}}}\frac{\overline{\omega }(r)}{{\Omega }}e^{-{\Phi }(r)}.
\end{eqnarray}
\\
 The \( \Omega _{c} \) is angular velocity in the approximation the amount
needed to produce shedding of mass at the star's equator, so that
this method of computation is not actually valid for this large a
value of \( \Omega  \).
\begin{figure}
{\par\centering
\resizebox*{10cm}{!}{\includegraphics{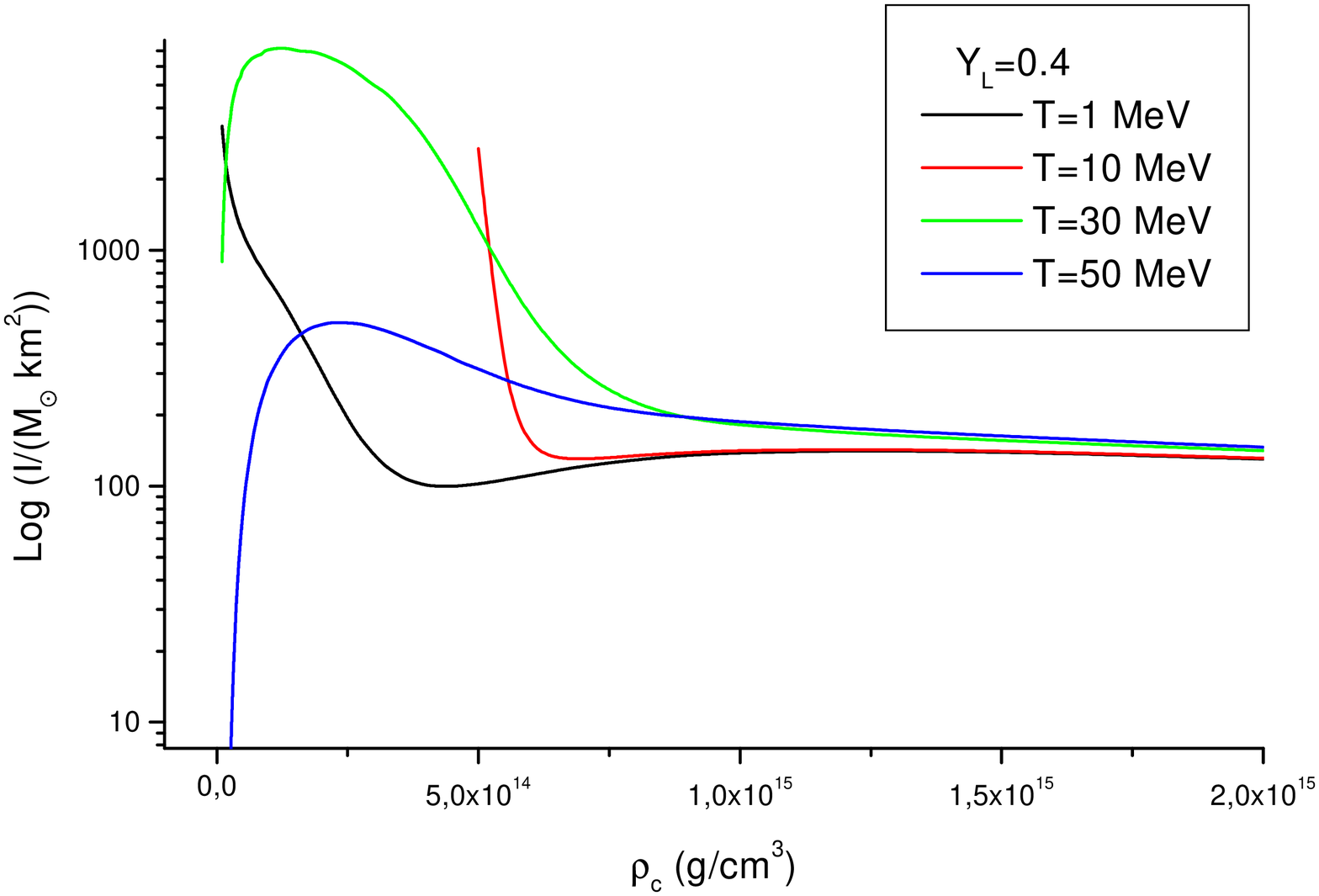}}
\par}
\caption{The moment of inertia \protect\( I\protect \), (in units
of $M_{\odot }$ $km^{2}$) as a function of the central density
for the different temperatures. The lepton number equals 0.4.}
\end{figure}
\begin{figure}
{\par\centering
\resizebox*{10cm}{!}{\includegraphics{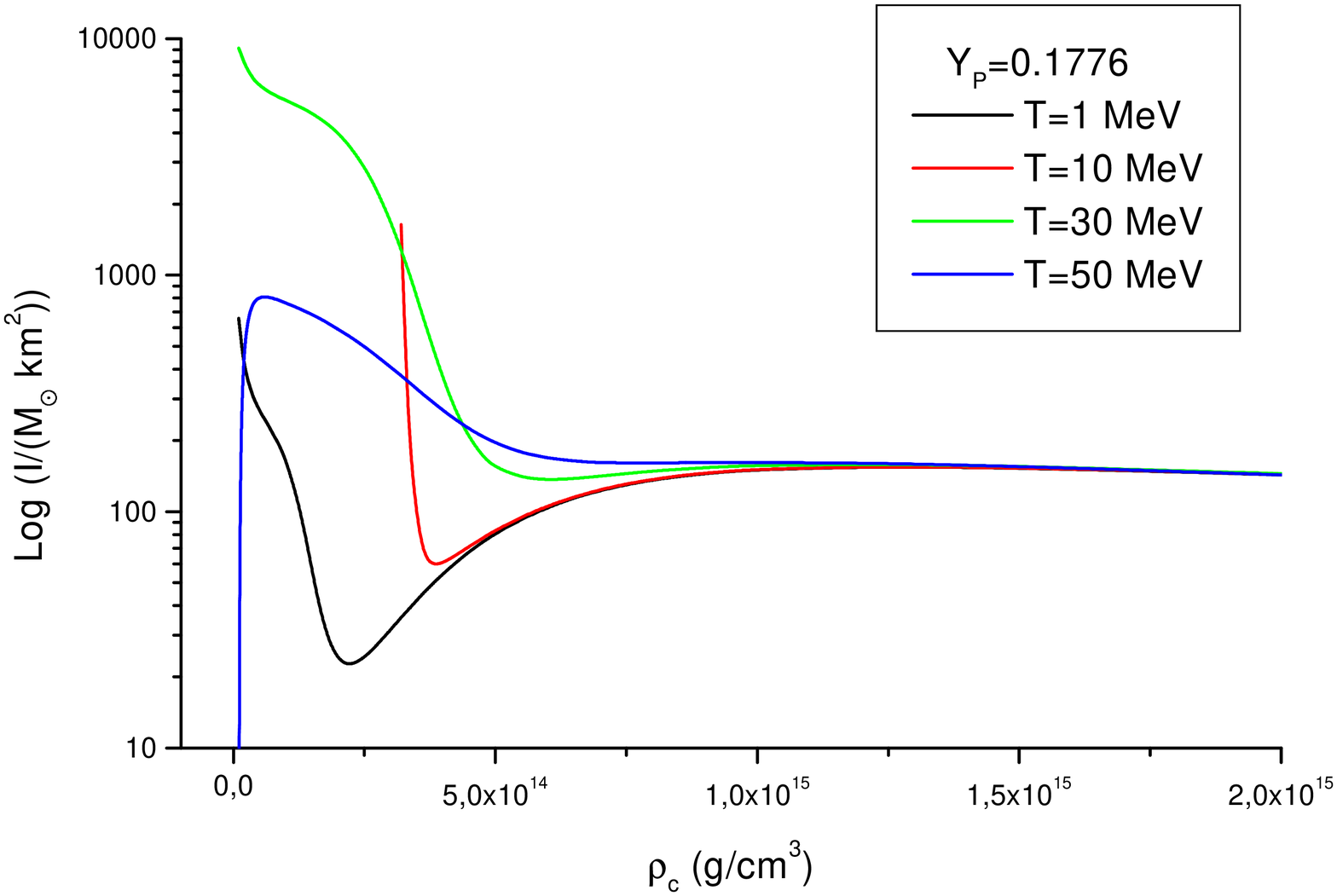}}
\par}
\caption{The moment of inertia \protect\( I\protect \), (in units
of $M_{\odot }$ $km^{2}$) as a function of the central density
for the different temperatures. The proton number equals 0.1776.}
\end{figure}
\begin{figure}
{\par\centering
\resizebox*{10cm}{!}{\includegraphics{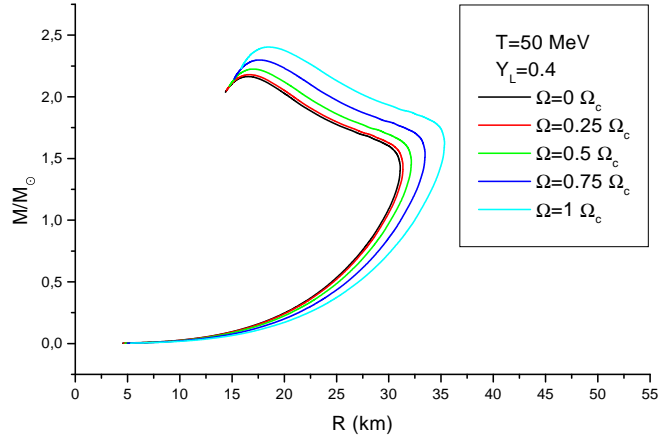}} \par}
\caption{The mass-radius relation for the protoneutron star for
different value of \protect\( \Omega\protect \). The temperature
$T$ and the lepton number $Y_{L}$ are fixed and equal $50$ $MeV$
and $0.4$, respectively.}
\end{figure}
\begin{figure}
{\par\centering
\resizebox*{10cm}{!}{\includegraphics{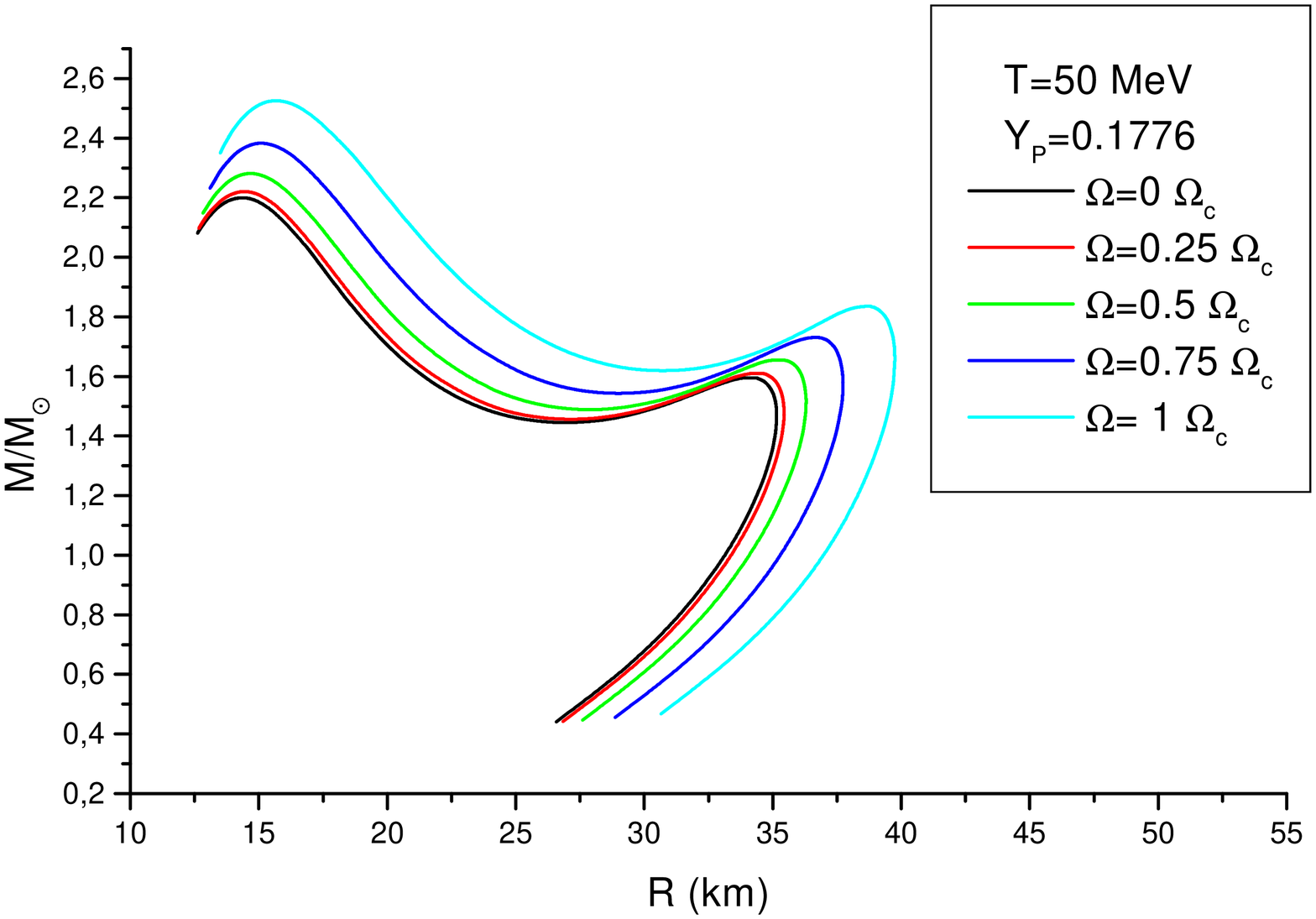}} \par}
\caption{The mass-radius relation for the protoneutron star for
different value of \protect\( \Omega\protect \). The temperature
$T$ and the proton fraction $Y_{P}$ are fixed and equal $50$ $MeV$
and $0.1776$, respectively.}
\end{figure}
\begin{figure}
{\par\centering
\resizebox*{10cm}{!}{\includegraphics{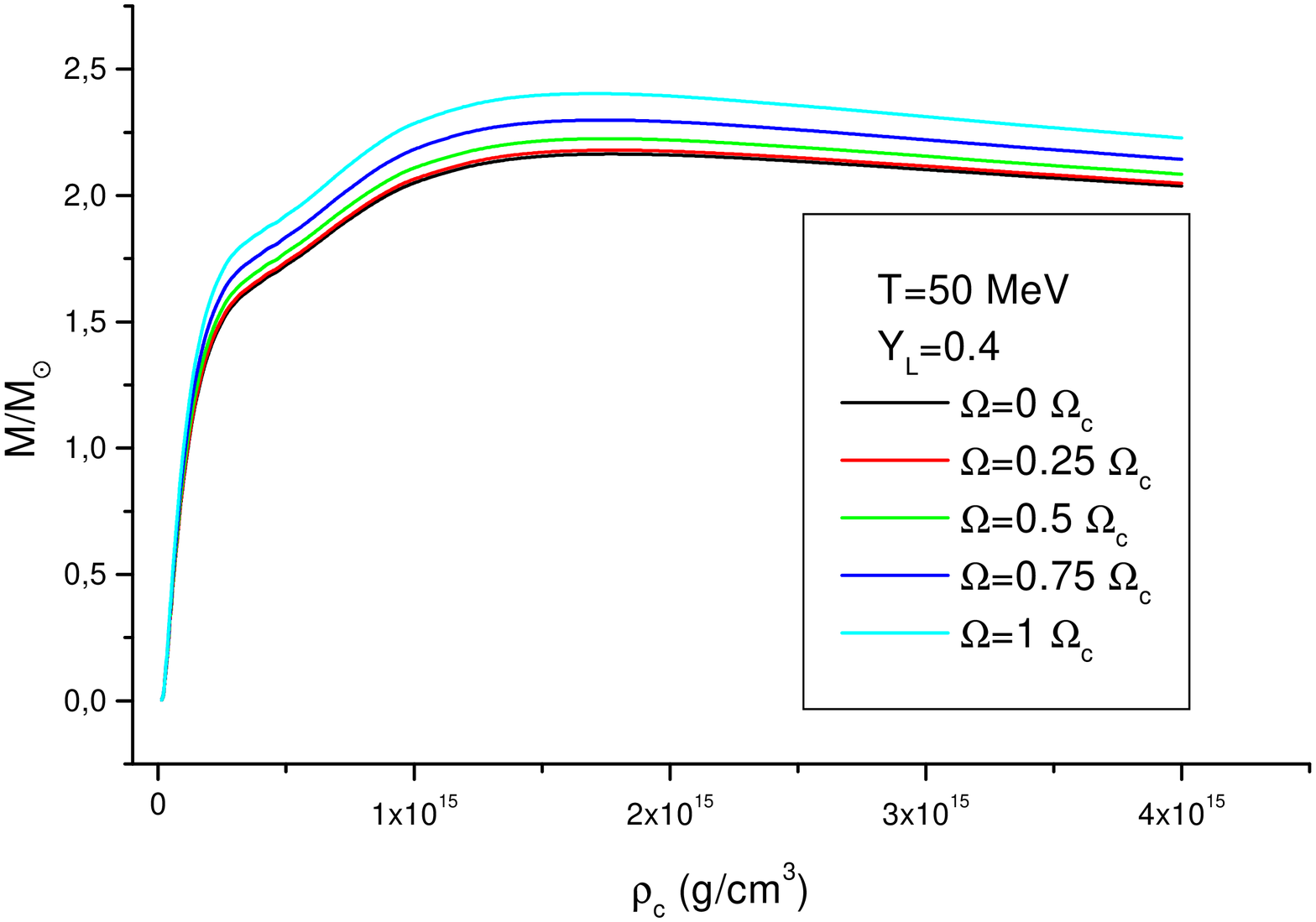}} \par}
\caption{The mass-central density relation for the protoneutron
star for different value of \protect\( \Omega\protect \). The
temperature $T$ and the lepton number $Y_{L}$ are fixed and equal
$50$ $MeV$ and $0.4$, respectively.}
\end{figure}
\begin{figure}
{\par\centering
\resizebox*{10cm}{!}{\includegraphics{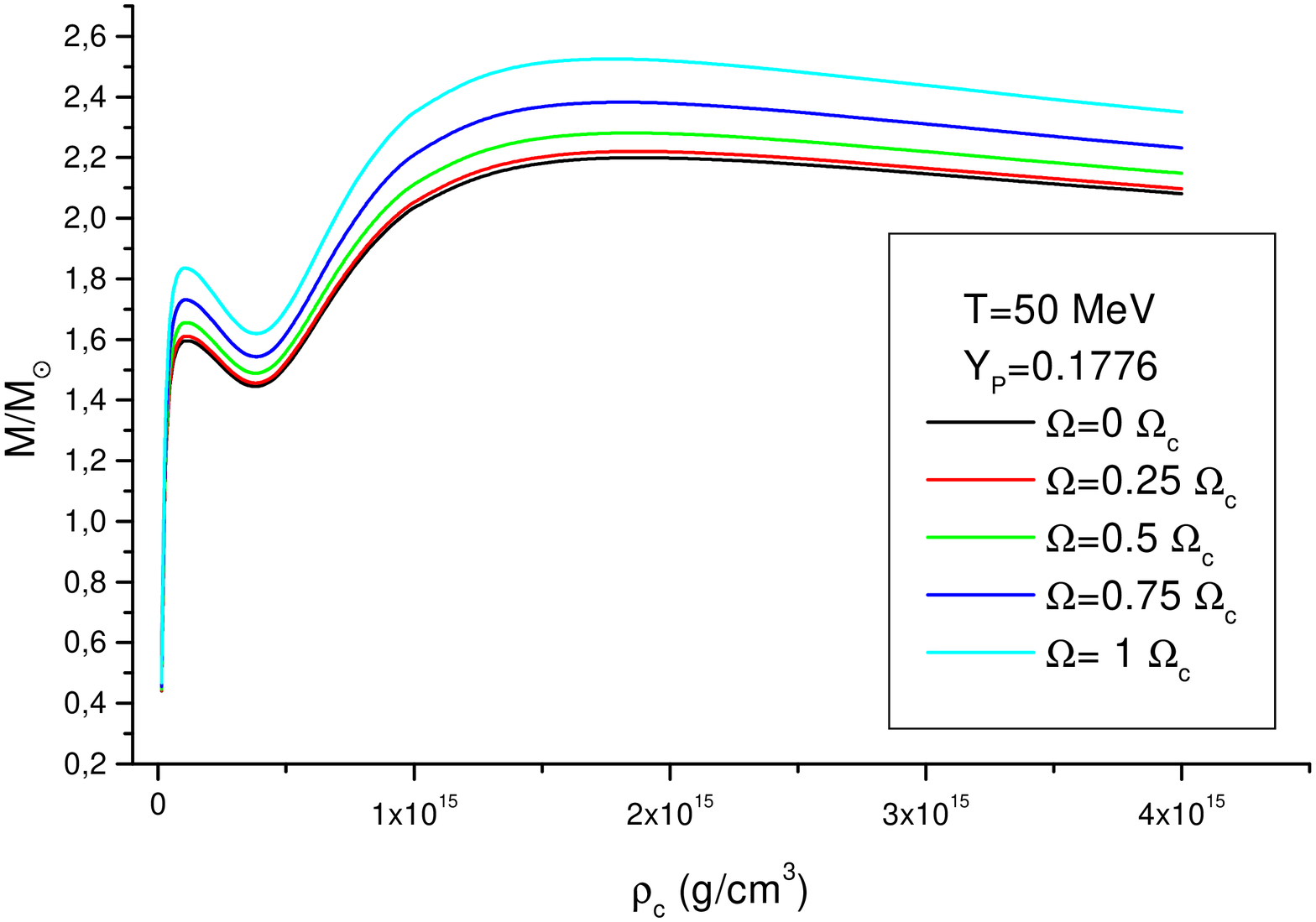}} \par}
\caption{The mass-central density relation for the protoneutron
star for different value of \protect\( \Omega\protect \). The
temperature $T$ and the proton fraction $Y_{P}$ are fixed and
equal $50$ $MeV$ and $0.1776$, respectively.}
\end{figure}
\section{The numerical results.}
In this paper the behaviour of a protoneutron star in the RMF
approach was examined. In Fig.1 the pressure P as a function of
the energy density is shown for the TM1 parameter set in the case
of finite temperature. The relevant temperature range is between
1 MeV up to 75 MeV. Figures 2 and 3 display the lepton fraction
$Y_{L}=(n_{e}+n_{\nu})/n_{B}$ and the entropy as a function of
central energy density. For different temperatures the proton
fraction $Y_{P}$ is fixed and equal $Y_{P}= 0.1776$. The pressure
as the function of the star radius is presented in Fig.4. The
total pressure of the protoneutron star being the sum of fermion
and meson parts is denoted by $P_{full}$. Particular contributions
$P_{\nu}$, $P_{p}$, $P_{n}$, $P_{e}$, $P_{out}$, $P_{o}$ mark the
pressure coming from neutrinos, protons, neutrons, electrons,
thermal plasma and pressure without neutrinos, respectively.
Neutrons pressure is the most significant component of the total
pressure whereas contributions coming from neutrinos and protons
are nearly the same in the star core. As the radius increases the
neutrino pressure starts to prevail over the remaining pressures.
Figs.5 and 6 depict the lepton number $Y_{L}$ and the entropy per
baryon $s$ as the function of the radius. The temperature is
fixed and equals $50$ MeV, $\rho_{c}= 1.8\times10^{15}$
$g/cm^{3}$. For the considerable value of the star radius the
lepton number and the entropy are almost constant whereas in the
outer layers of the protoneutron star both of them grow very
steeply. The mass-radius relations for protoneutron stars for
different temperature cases and constant lepton $Y_{L}$ and
proton $Y_{P}$ number is presented in the next two figures. The
stable and unstable areas of the stellar configuration (the
dotted line) are presented. The possible stellar evolution tracks
with the constant baryon number are pointed by arrows. Considering
configurations obtained for different temperatures one can see
the possible evolution tracks. They are presented in Table
\ref{tab8} and Table \ref{tab9}.
\begin{table}
\begin{tabular}{|c|c|c|}
\hline \( T\, [\, MeV\, ] \)&
 \( R\, [\, km\, ] \)&
 \( M\, [\, M_{\odot }\, ] \)\\
\hline \( 50 \)&
 \( 20.8981 \)&
 \( 1.9751 \)\\
 \hline \( 30 \)&
 \( 79.9505 \)&
 \( 3.2871\)\\
  \(  \)&
 \( 40.6529\)&
 \( 1.9139 \)\\
\hline \( 10 \)&
 \( 40.5229\)&
 \( 1.6566\)\\
\hline \( 1 \)&
 \( 20.2331 \)&
 \( 1.6328 \)\\
\hline
\end{tabular}
\caption{\label{tab8}The evolution tracks for \protect\(
Y_{L}=0.4\protect \).}
\end{table}
\begin{table}
\begin{tabular}{|c|c|c|}
\hline \( T\, [\, MeV\, ] \)&
 \( R\, [\, km\, ] \)&
 \( M\, [\, M_{\odot }\, ] \)\\
\hline \( 50 \)&
 \( 22.1870 \)&
 \( 1.5560 \)\\
 \hline \( 30 \)&
 \( 84.6644 \)&
 \( 3.1236\)\\
  \(  \)&
 \( 32.2714\)&
 \( 1.3731 \)\\
\hline \( 10 \)&
 \( 25.2496\)&
 \( 1.2360\)\\
\hline \( 1 \)&
 \( 15.8983 \)&
 \( 1.2232 \)\\
\hline
\end{tabular}
\caption{\label{tab9}The evolution tracks for \protect\(
Y_{P}=0.1776\protect \).}
\end{table}
At the initial time $t=0$ s the protoneutron star is characterized
by the barion mass of the value of 1 $M_{\odot}$ and the proton
fraction $Y_{P}=0.2$, the temperature equals 30 MeV. In
accordance with the radial profiles of temperature present in
paper \cite{toki} after 15 s the temperature and the proton
fraction drop to 10 MeV and $1/9$, respectively.
 For comparison the mass-radius dependence for the protoneutron
star with constant lepton and proton number and in the
temperature equal $30$ MeV are presented in Fig.9. The unstable
configurations appear and they are marked by dotted lines. The
same relation is shown for the star in zero temperature limit and
with zero neutrino chemical potential (without neutrinos), than
the proton concentration is constrained by the $\beta$
equilibrium. The mass-central density ($\rho_{c}$) functions for
different temperatures and fixed lepton number $Y_{L}$ (Fig.11)
and proton fraction $Y_{P}$ (Fig.12) are shown on Figs.11 and 12.
From relations presented on Figs.7 and 8 one can see the density
ranges for which the stars are stable. Using the results of
Figs.7 and 8 the density ranges of unstable protoneutron stars
configurations can be estimated. The moment of inertia ($I$) of a
protoneutron star versus the central density ($\rho_{c}$) for
different values of temperature are presented. Figures 15 and 16
show the subsequence mass-radius relations of a hot protoneutron
star ($T=50$ $MeV$) obtained for different angular velocities.
The lepton and proton number are constant. The greatest values of
the angular velocity the more massive and bigger are the stars.
Figures 17 and 18 compare the mass-central density relations for
protoneutron stars with fixed values of lepton $Y_{L}$ and proton
$Y_{P}$ number. The temperature equals $50$ $MeV$ obtained for
different angular velocities. In Fig.18 the density range of
stable rotating protoneutron star is visable.
 The main numerical results is collected in the Table
\ref{tab: 3} and the Table \ref{tab: 4}, where \( Y_{L}=0.4 \) and
\( \rho _{c}=5 \) \( 10^{14} \) \( g/cm^{3} \).
\begin{table}
\begin{tabular}{|c|c|c|c|c|}
\hline
\( T\, [\, MeV\, ] \)&
 \( R\, [\, km\, ] \)&
 \( M\, [\, M_{\odot }\, ] \)&
 \( M_{B}\, [\, M_{\odot }\, ] \)&
 \( I\, [\, M_{\odot }\, km^{2}] \)\\
\hline \( 1 \)&
 \( 14.3886 \)&
 \( 2.1312 \)&
 \( 2.3057 \)&
 \( 130.4967 \)\\
 \hline \( 10 \)&
 \( 18.3521 \)&
 \( 2.135\)&
 \( 2.3051 \)&
 \( 131.3594 \)\\
\hline
\( 30 \)&
 \( 80.0136 \)&
 \( 3.3723 \)&
 \( 1.7768 \)&
 \( 123.7578 \)\\
 \(  \)&
 \( 19.6225 \)&
 \( 2.1694 \)&
 \( 2.2806 \)&
 \( 146.7724 \)\\
\hline
\( 50 \)&
 \( 16.4648 \)&
 \( 2.1659 \)&
 \( 2.1226 \)&
 \( 152.4982 \)\\
\hline
\( 75 \)&
 \( 12.5894 \)&
 \( 1.9839 \)&
 \( 1.7173 \)&
 \( 116.9361 \) \\
\hline\end{tabular} \caption{\label{tab: 3}The mass, the radius
and the moment of inertia for \protect\( Y_{L}=0.4\protect \).}
\end{table}
\begin{table}
\begin{tabular}{|c|c|c|c|c|c|}
\hline
\( T\, [\, MeV\, ] \)&
 \( R\, [\, km\, ] \)&
 \( \overline{R}\, [\, km\, ] \)&
 \( M\, [\, M_{\odot }\, ] \)&
 \( \overline{M}\, [\, M_{\odot }\, ] \)&
 \( J\, [\, M_{\odot }\, km^{2}\, s^{-1}\, ] \)\\
\hline \( 1\)&
 \( 14.3886 \)&
 \( 15.6158 \)&
 \( 2.1312\)&
 \( 2.4281 \)&
 \( 1271590 \)\\
 \hline \( 10\)&
 \( 18.3521\)&
 \( 20.9522 \)&
 \( 2.135\)&
 \( 2.2798 \)&
 \( 889378 \)\\
\hline
\( 30 \)&
 \( 80.0136 \)&
 \( 92.0202 \)&
 \( 3.3722 \)&
 \( 3.7851 \)&
 \( 6658570 \)\\
 \(  \)&
 \( 19.6225 \)&
 \( 22.5576 \)&
 \( 2.1694 \)&
 \( 2.3129 \)&
 \( 906022\)\\
\hline
\( 50 \)&
 \( 16.4648 \)&
 \( 18.2002 \)&
 \( 2.1659 \)&
 \( 2.4031 \)&
 \( 1223790 \)\\
\hline
\( 75 \)&
 \( 12.5894 \)&
 \( 13.4091 \)&
 \( 1.9839 \)&
 \( 2.3060 \)&
 \( 1343250 \) \\
\hline
\end{tabular}
\caption{\label{tab: 4}The mass, the radius and the angular
momentum for \protect\( Y_{L}=0.4\protect \).}
\end{table}
\section{The conclusion.}
The main aim of this paper was to study the protoneutron star
parameters especially the masses and radii as the most sensitive
ones to the form the equation of state. The employed form of EOS
in the TM1 set which was extended to the finite temperature
cases. The considered model comprises not only nucleons and
electrons, which are necessary for the $\beta$ stable matter, but
the neutrinos as well. The presence of neutrinos is characteristic
for hot young protoneutron stars. Neutrinos give not negligible
contribution to the pressure and energy density. As the radius of
protoneutron stars increases the neutrino pressure starts to
prevail over the pressure coming from the remaining components.
Constructing the mass-radius relations for protoneutron stars the
stable and unstable areas appeared. This indicate that there are
several possible stellar evolution tracks for the configuration
with constant baryon number. These evolutionary tracks are
strictly connected with the decreasing star temperature and thus
with the decreasing neutrino chemical potential and lepton
number. The final result is the cold deleptonized (the neutrino
chemical potential equals zero) object. The influence of rotation
on the protoneutron star parameters is significant. In this paper
the case of slowly rotating protoneutron star is considered. The
obtained unstable configuration for the fixed temperature ($T=30$
MeV) and lepton number ($Y_{L}=0.4$) occurring in the density
range ($1.4-6.5\times 10^{14}$ $g/cm^{3}$) are relevant to rapid
change in the moment of inertia of the star (Fig.13 and Fig.14).
For the rotating star the mass-radius relation depends on the star
angular velocity thus the values of $\Omega$ affects the
instability areas. For the employed EOS there exists the ranges
of masses where hot, lepton rich protoneutron star can be
stabilized against gravity but loosing leptons these objects
became gravitationally unstable. The instability areas as
strictly connected with the presence of neutrinos and the thermal
pressure effects.

\end{document}